\documentclass[pra,aps,nofootinbib,showpacs,floatfix,superscriptaddress,amsmath,twocolumn]{revtex4-1}

\usepackage{graphicx}
\usepackage{amsmath}
\usepackage{amssymb}
\usepackage{color}
\usepackage{bm}
\usepackage{hyperref}
\hypersetup{colorlinks,allcolors=black}
\usepackage{url}

\newcommand {\fabs}[1] {\left| #1 \right|}
\newcommand {\fnorm}[1] {\left\Vert #1 \right\Vert}

\newcommand {\fabsq}[1] {\left\vert #1 \right\vert^2}

\newcommand{\ket}[1]{\ensuremath{|#1\rangle}}

\newcommand{\ketbra}[2]{|#1\rangle\langle#2|}
\newcommand{\braXket}[3]{\langle#1|#2|#3\rangle}

\newcommand{\cF}{{\cal{F}}}

\newcommand{\Eqref}[1]{Eq. \eqref{#1}}

\newcommand{\hess}[1]{\text{Hess}\left[#1\right]}

\begin{document}
\title{Improved anharmonic trap expansion through enhanced shortcuts to adiabaticity}
\author{C. Whitty}
\email{c.whitty@umail.ucc.ie}
\affiliation{Department of Physics, University College Cork, Cork, Ireland}

\author{A. Kiely}
\affiliation{School of Physics, University College Dublin, Belfield, Dublin 4, Ireland}
\affiliation{Centre for Quantum Engineering, Science, and Technology, University College Dublin, Belfield, Dublin 4, Ireland}

\author{A. Ruschhaupt}
\affiliation{Department of Physics, University College Cork, Cork, Ireland}

\begin{abstract}%%%%%%%%%%%%%%%%%%%%%%%%%%%%%%%%%%%%%%%%%%%%%%%%%%%%%%%%%%%%%%%%%%%%%%%
Shortcuts to adiabaticity (STA) have been successfully applied both theoretically and experimentally to a wide variety of quantum control tasks.
In previous work the authors have developed an analytic extension to shortcuts to adiabaticity, called enhanced shortcuts to adiabaticity (eSTA), that extends STA methods to systems where STA cannot be applied directly [Phys. Rev. Research 2, 023360 (2020)].
Here we generalize this approach and construct an alternative eSTA method that takes advantage of higher order terms.
We apply this eSTA method to the expansion of both a Gaussian trap and accordion lattice potential, demonstrating the improved fidelity and robustness of eSTA.
\end{abstract}
\maketitle

%%%%%%%%%%%%%%%%%%%%%%%%%%%%%%%%%%%%%%%%%%%%%%%%%%%%%%%%%%%%%%%%%%%%%%%
\section{Introduction}
%%%%%%%%%%%%%%%%%%%%%%%%%%%%%%%%%%%%%%%%%%%%%%%%%%%%%%%%%%%%%%%%%%%%%%%

Fast high-fidelity control of quantum systems is a key requirement for the implementation of future quantum technologies \cite{glaserTrainingSchrodingerCat2015b}.
Specifically, analytical control solutions are particularly desirable as they are simpler, provide greater physical insight and allow for additional stability requirements \cite{ruschhauptOptimallyRobustShortcuts2012a,kielyFastStableManipulation2015b}.
Shortcuts to Adiabaticity (STA) are a broad class of analytical control techniques that mimic adiabatic evolution on much shorter timescales \cite{chenFastOptimalFrictionless2010b,torronteguiChapterShortcutsAdiabaticity2013a,guery-odelinShortcutsAdiabaticityConcepts2019b,campoFocusShortcutsAdiabaticity2019}.
STA have been applied in many different contexts; engineering quantum heat engines \cite{campoMoreBangYour2014,delCampo2018,liEfficientNonlinearFeshbach2018b}, creating exotic angular momentum states in optical lattices \cite{kielyShakenNotStirred2016}, designing experimentally realizable fast driving of many-body spin systems \cite{saberiAdiabaticTrackingQuantum2014b}, speed up STIRAP population transfer \cite{demirplakAdiabaticPopulationTransfer2003,bergmannPerspectiveStimulatedRaman2015,masudaFastForwardAssistedSTIRAP2015,vitanovStimulatedRamanAdiabatic2017a}, and to inhibit unwanted transitions in two and three level systems \cite{kielyInhibitingUnwantedTransitions2014}.

However, STA methods can have limitations.
Some STA techniques could require non-trivial physical implementation (e.g. counterdiabatic driving), while other STA techniques may only be easily applied to small or highly symmetrical systems (e.g. Lewis-Riesenfeld invariants)\cite{guery-odelinShortcutsAdiabaticityConcepts2019b, torronteguiChapterShortcutsAdiabaticity2013a}.
This difficulty motivated the development of Enhanced Shortcuts to Adiabaticity (eSTA) where STA solutions can be perturbatively corrected to perform well on more complex quantum systems \cite{whittyQuantumControlEnhanced2020}.
This method is broadly applicable, since many applications of STA techniques have already used idealized Hamiltonian descriptions e.g. single effective particles \cite{ julia-diazFastGenerationSpinsqueezed2012, kielyFastRobustMagnon2021}, few-level descriptions \cite{kielyShakenNotStirred2016, martinez-garaotVibrationalModeMultiplexing2013,bensenySpatialNonadiabaticPassage2017a,liQubitGatesSimultaneous2018a}, and mean field Hamiltonians \cite{takahashiShortcutsAdiabaticityQuantum2017}.
Indeed, eSTA has been applied to the transport of neutral atoms in optical conveyor belts \cite{hauckSingleatomTransportOptical2021,hauckCoherentAtomTransport2021a}, and additionally has been shown to have intrinsic robustness \cite{whittyRobustnessEnhancedShortcuts2022}.

In this work we generalize the original eSTA approach and formulate an alternative eSTA scheme.
While the original eSTA scheme uses the assumption that perfect fidelity can be achieved near the starting STA scheme, the alternative eSTA scheme does not require this assumption by using higher order terms.
We show how these higher order terms can be systematically calculated using time-dependent perturbation theory.

We apply the original and alternative eSTA schemes to atomic trap expansion, using the physical settings of an optical dipole trap and an optical accordion lattice.
Trap expansion of anharmonic potentials using STA has been studied \cite{torronteguiFastTransitionlessExpansion2012, luFastTransitionlessExpansions2014a}, and faster than adiabatic trap expansion has been experimentally realized using STA invariant-based engineering, in a cold atomic cloud \cite{schaffFastOptimalTransition2010}, one dimensional Bose gas \cite{rohringerNonequilibriumScaleInvariance2015}, a Fermi gas \cite{dengShortcutsAdiabaticityStrongly2018,dengSuperadiabaticQuantumFriction2018} and loading a Bose Einstein condensate (BEC) into an optical lattice \cite{masudaHighFidelityRapidGroundState2014a,zhouShortcutLoadingBose2018}.
The dynamic control of lattice spacing in optical accordions is another important trap expansion setting \cite{williamsDynamicOpticalLattices2008,liRealtimeControlPeriodicity2008,taoWavelengthlimitedOpticalAccordion2018}.
There have been a variety of experimental realizations of optical accordions;
dynamically expanding the lattice spacing of an optical accordion loaded with ultra-cold atoms \cite{al-assamUltracoldAtomsOptical2010}, expansion of a one dimensional BEC \cite{fallaniBoseEinsteinCondensateOptical2005a} and loading and compression of a two dimensional tunable Bose gas in an optical accordion \cite{villeLoadingCompressionSingle2017}.

This paper is organized as follows.
In Sec. \ref{sect_eSTA_formalism} we introduce the generalized eSTA formalism.
Then we look at trap expansion in Sec. \ref{sect_trap_opening} and compare STA and eSTA control schemes, considering their sensitivity to amplitude noise in both trap models.

%%%%%%%%%%%%%%%%%%%%%%%%%%%%%%%%%%%%%%%%%%%%%%%%%%%%%%%%%%%%%%%%%%%%%%%
\section{Generalized eSTA formalism \label{sect_eSTA_formalism}}
%%%%%%%%%%%%%%%%%%%%%%%%%%%%%%%%%%%%%%%%%%%%%%%%%%%%%%%%%%%%%%%%%%%%%%%

In the following we give a generalized derivation of eSTA, complementary to the original formalism outlined in \cite{whittyQuantumControlEnhanced2020}, which allows the formulation of an alternate eSTA scheme.

The goal of eSTA is to control a quantum system with Hamiltonian $H_{s}$.
Specifically, we want to evolve the initial state $\ket{ \Psi_0}$ at time $t=0$ to the target state $\ket{ \Psi_T}$ in a given total time $t_f$.
We assume that $H_{s}$ can be approximated by an existing Hamiltonian $H_0$ with known STA solutions, that we refer to as the idealized STA system.
In detail, we assume that there exists a parameter $\mu$ that varies continuously from $\mu_s$ to $0$, such that $H_s=H_{\mu_s}$ approaches $H_0$ as $\mu$ approaches $0$.
In later examples $\mu$ will represent the anharmonicity of the experimental trapping potential, with the idealized STA system taking the form of a time-dependent harmonic oscillator.
We parameterize the control scheme of $H_{\mu_s}$ by a vector $\vec{\lambda}$, which represents the deviation from the original STA control scheme ($\vec{\lambda}=\vec{0}$).
Our objective is to derive a correction to the STA scheme that improves the fidelity, which we label $\vec{\lambda}_s$.

There are two main steps behind the derivation of eSTA. 
The first is the assumption that $\mu$ and $\| \vec{\lambda}\|$ are small such that the fidelity landscape around $(\mu=0, \vec{\lambda}=\vec{0})$ can be well approximated to second order in $\mu$ and $\vec{\lambda}$.
Secondly, we can take advantage of the known time evolution operator of the STA system to derive an improved control scheme analytically using time-dependent perturbation theory.

\subsection{eSTA construction}
Throughout the following derivation of eSTA, we assume that at the initial and target states of $H_{\mu_s}$ can be approximated by the known eigenstates of $H_0$ at initial and final times.
We define the fidelity
\begin{align}
F(\mu,\vec{\lambda})=\fabs{\braXket{\Psi_T}{U_{\mu,\vec{\lambda}}(t_f,0)}{\Psi_0}}^2,
\end{align}
where the time evolution is explicitly parameterized by $\mu$ and $\vec{\lambda}$ through the Hamiltonian $H_\mu (\vec{\lambda},t)$.

For a given $H_{\mu_s}$ we derive the eSTA control vector $\vec{\lambda}_s$ by approximating several quantities that allow us to construct a parabola in $\vec{\lambda}$ for fixed $\mu=\mu_s$.
This parabola projects a path of increasing fidelity, and using the eSTA formalism we calculate the $\vec{\lambda}_s$ that corresponds to the peak of this parabola.
To illustrate this construction explicitly, we let $\vec{\lambda} = \epsilon \, \hat{v}$ and set
\begin{align}
f(\epsilon) = F(\mu_s,\epsilon \, \hat{v}),
\end{align}
with $ \hat{v} = \nabla_{\vec{\lambda}} F(\mu_s, \vec{0}) / \fnorm{\nabla_{\vec{\lambda}} F(\mu_s, \vec{0})}$.
We now approximate
\begin{align}\label{eq:eSTA_parabola}
f(\epsilon) \approx f(0)+ \epsilon f'(0) + \frac{\epsilon^2}{2} f''(0),
\end{align}
with
\begin{align}\label{eq:eSTA_equs}
f(0) &= F(\mu_s, \vec{0}),
\nonumber \\
f'(0) &= \fnorm{\nabla_{\vec{\lambda}} F(\mu_s, \vec{0})},
\nonumber \\
f''(0) &=\hat{v}^{T} \:\hess{ F(\mu_s, \vec{0})} \hat{v},
\end{align}
where $\hess{ F(\mu_s, \vec{0})}$ is the Hessian matrix of second order partial derivatives of $F$ with respect to the components of $\vec{\lambda}$, and the superscript $T$ denotes vector transposition.

In the original eSTA approach \cite{whittyQuantumControlEnhanced2020}, the parabola is constructed using approximations to the fidelity $F(\mu_s, \vec{0})$, the gradient $\nabla_{\vec{\lambda}} F(\mu_s, \vec{0})$, together with the assumption that the optimal control vector $\vec{\lambda}_{s}^{(1)}$ can achieve perfect fidelity i.e. $F(\mu_s, \vec{\lambda}_{s}^{(1)}) = 1$.
This leads to $\epsilon_{s}^{(1)} = 2[1-f(0)]/f'(0)$, with
\begin{align}\label{eq:lambda_orig}
\vec{\lambda}_{s}^{(1)} = \frac{2\left[ 1 - F(\mu_s, \vec{0}) \right]}{\fnorm{\nabla_{\vec{\lambda}} F(\mu_s, \vec{0})}} \: \hat{v}.
\end{align}
We label this original method eSTA$_1$, and note that it does not use approximations to terms beyond the gradient.

Using the generalized derivation presented later in Sec. \ref{sec:eSTA_derive}, we can obtain a simple approximation to the second order term $f^{''}(0)$ in \Eqref{eq:eSTA_equs}.
Using this higher order term we derive an alternative eSTA scheme that we label $\text{eSTA}_2$.
We construct this scheme by noting that the maximum of $f(\epsilon)$ will be at $\epsilon_{s}^{(2)} = -f'(0)/f''(0)$, and the $\text{eSTA}_2$ control vector now takes the form
\begin{align}\label{eq:lambda_Hessian}
\vec{\lambda}_{s}^{(2)} = -\frac{\nabla_{\vec{\lambda}} F(\mu_s, \vec{0})}{\hat{v}^{T} \:  \hess{  F(\mu_s, \vec{0}) }\hat{v}}.
\end{align}

\begin{figure}[t]
\begin{center}
\includegraphics[width=0.99\linewidth]{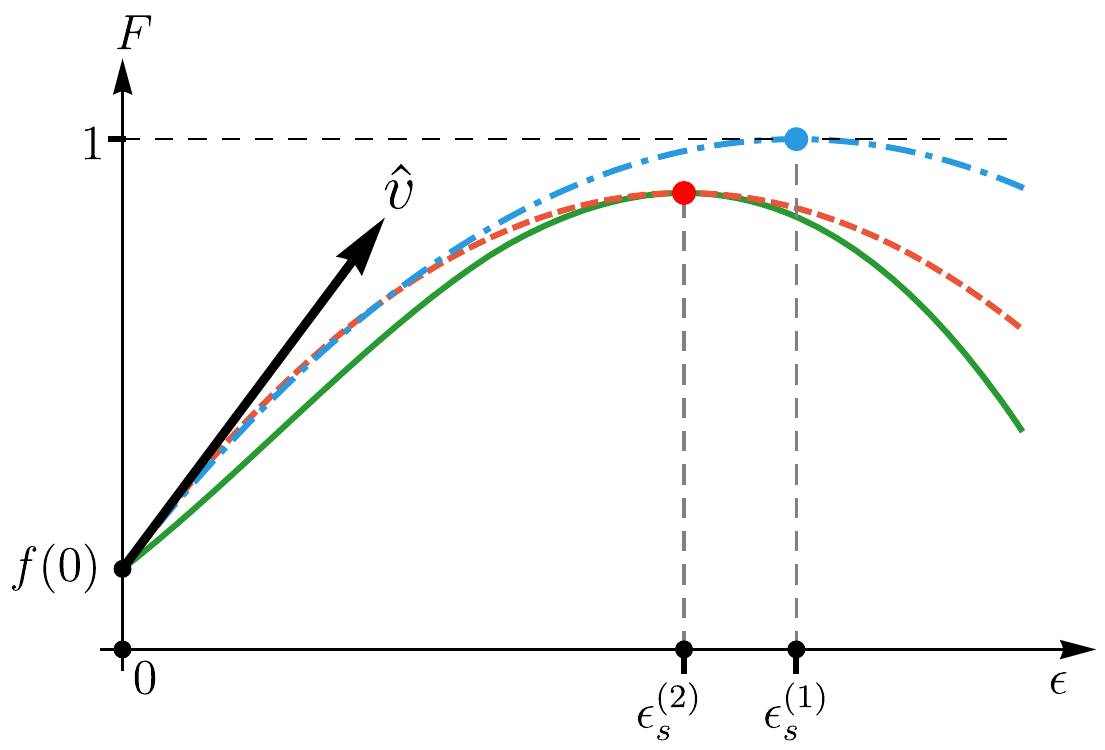}
\end{center}
\caption{\label{fig:1_eSTA_schematic}(color online)
Diagram of eSTA construction, $F(\mu,  \epsilon \hat{v})$ vs. $\epsilon$.
The true fidelity landscape (solid green), $\text{eSTA}_1$ parabolic approximation (dot-dashed blue) and $\text{eSTA}_2$ parabola (dashed red) are shown.
The normalized gradient $\hat{v}$ is represented by the solid black arrow, and $F(\mu_s,\vec{0})$ corresponds with $f(0)$.
The result of applying $\text{eSTA}_2$ is the improved control vector $\vec{\lambda}_s^{(2)}= \epsilon_s^{(2)} \hat{v}$, which is shown matching the peak of the true fidelity landscape well.
}
\end{figure}

We schematically represent the parabola construction of \Eqref{eq:eSTA_parabola} in Fig. \ref{fig:1_eSTA_schematic}.
Note that $\text{eSTA}_1$ (dot-dashed blue line) can overshoot the desired optimal $\vec{\lambda}_s$, due to the assumption that $F=1$ at $\vec{\lambda}_{s}^{(1)}$.
At the expense of calculating the Hessian term in \Eqref{eq:eSTA_equs} this assumption can be removed.
Note that in later examples of trap expansions we will show quantitative versions of the schematic in Fig. \ref{fig:1_eSTA_schematic}.

%**********************************************************************
%**********************************************************************

\subsection{Perturbative approximations for eSTA control}\label{sec:eSTA_derive}

To calculate the eSTA control vector $\vec{\lambda}_s$ for either $\text{eSTA}_1$ or $\text{eSTA}_2$, we approximate the quantities in \Eqref{eq:eSTA_equs} using the known solutions to the STA system $H_0$ (which can be obtained for example by using invariant-based inverse engineering \cite{chenFastOptimalFrictionless2010b,torronteguiChapterShortcutsAdiabaticity2013a,guery-odelinShortcutsAdiabaticityConcepts2019b}) and time-dependent perturbation theory.
We label the STA solutions $\{\ket{\chi_n (t)}\}$, with $\ket{\chi_n (t)} = U_{0,\vec{0}} (t,0) \ket{\chi_n (0)}$.
The time evolution operator of the STA system can be written as
\begin{align}\label{eq:H0_time_evolu}
U_{0,\vec{0}} (t,s) =\sum_{n=0}^{\infty} \ketbra{\chi_n(t)}{\chi_n(s)}.
\end{align}
The time evolution of a general $H_{\mu}$ can be expanded using time-dependent perturbation theory as
\begin{align}
U_{\mu,\vec \lambda} (t,s) = U_{0,\vec{0}} (t,s) + \sum_{n=1}^\infty  U_{\mu,\vec{\lambda}}^{(n)}(t,s),
\end{align}
where
\begin{align}
U_{\mu,\vec{\lambda}}^{(n)}(t_f,0) = \left( - \frac{i}{\hbar} \right)^n 
& \int^{(n)}  
U_{0,\vec{0}} \left( t_f,t_n \right) \Delta H (t_n) \dots
\nonumber \\
& \dots U_{0,\vec{0}} \left( t_2,t_1 \right) \Delta H (t_1) U_{0,\vec{0}} \left( t_1,0 \right),
\end{align}
and $\Delta H_{\mu}(\vec{\lambda},t) = H_{\mu} (\vec{\lambda},t) - H_0 (\vec{0},t)$, with the multi-integrals defined using the notation
\begin{align}
\int^{(n)} \equiv \int_0^{t_f} dt_n \int_0^{t_n} dt_{n-1} \dots \int_0^{t_3} dt_{2} \int_0^{t_2} dt_{1}.
\end{align}
We now set
\begin{align}
\cF &= \braXket{\chi_0(t_f)}{U_{\mu,\vec{\lambda}}(t_f,0)}{\chi_0(0)}
\nonumber \\
&= \sum_{n=0}^\infty \braXket{\chi_0(t_f)}{U^{(n)}(t_f,0)}{\chi_0(0)}
\nonumber \\
&= 1 + \sum_{n=1}^\infty \cF^{(n)},
\end{align}
where we have used $\braXket{\chi_0(t_f)}{U_{0,\vec{0}}(t_f,0)}{\chi_0(0)} = 1$.
% (by definition of the STA system).
Thus the fidelity becomes
\begin{align}\label{eq:fid_expand_exact}
F(\mu, \vec{\lambda}) &= \fabs{\cF(\mu, \vec{\lambda})}^2
\nonumber \\
=&
\left[ 1+ \sum_{n=1}^{\infty} \cF^{(n)} \right]
\left[ 1+ \sum_{m=1}^{\infty} \left(\cF^{(m)}\right)^* \right]
\nonumber \\
=& 1 + 
2 \sum_{n=1}^{\infty} \text{Re} \left[ \cF^{(n)} \right]
+ \sum_{n=1}^{\infty} \sum_{m=1}^{\infty} \text{Re} \left[ \cF^{(n)} \left(\cF^{(m)}\right)^* \right].
\end{align}
Now we define
\begin{align}
\Gamma_{n,m}(t) = \braXket{\chi_n(t)}{\Delta H_{\mu}(\vec{\lambda},t)}{\chi_m(t)},
\end{align}
and by repeated use of \Eqref{eq:H0_time_evolu} we have
\begin{align}\label{eq:Fn}
\cF^{(n)} = \left(\frac{-i}{\hbar}\right)^n
\sum_{m_1=0}^\infty \cdots \sum_{m_{n-1}=0}^\infty
\int^{(n)}
\prod_{l=0}^{n-1} \Gamma_{m_{l+1},m_l} \left( t_{l+1} \right),
\end{align}
where $m_0=m_n=0$ and the factors in the product commute.

The advantage of this notation is that $ \cF^{(n)}$ breaks the time interval into $n$ nested integrals, and we can collect the integrals up to second order and obtain
\begin{align}
\cF^{(1)} =& -\frac{i}{\hbar} \int_0^{t_f} \: dt_1 \Gamma_{0,0} (t_1) ,
\nonumber \\
\cF^{(2)} =& -\frac{1}{\hbar^2} \int_0^{t_f} dt_2 \int_0^{t_2} dt_1 \sum_{m=0}^{\infty} \Gamma_{0,m} (t_2) \Gamma_{m,0} (t_1).
\end{align}
If we consider the fidelity up to double integrals (i.e. $n=2$), we can write
\begin{align}\label{eq:fid_expansion}
F \left( \mu, \vec{\lambda} \right) &\approx 1 + \frac{1}{\hbar^2} \fabs{\int_0^{t_f} dt \: \Gamma_{0,0} (t)}^2
\nonumber \\
-& \frac{2}{\hbar^2} 
\int_0^{t_f} dt_2 \int_0^{t_2} 
\sum_{m=0}^{\infty} \text{Re} \left[ \Gamma_{m,0}^* (t_2) \Gamma_{m,0} (t_1) \right],
\end{align}
which can be simplified to
\begin{align}\label{eq:eSTA_fid_new}
F \left( \mu, \vec{\lambda} \right) \approx 1 - \frac{1}{\hbar^2}
\sum_{m=1}^{\infty} \fabs{ \int_0^{t_f} dt \: \Gamma_{m,0} (t) }^2.
\end{align}
We derive the gradient approximation by directly taking the derivative of \Eqref{eq:eSTA_fid_new},
\begin{align}\label{eq:grad_approx_new}
\frac{\partial F}{\partial \lambda_k } \approx
-\frac{1}{\hbar^2}
\sum_{m=1}^{\infty} 
\Bigg[&
\int_0^{t_f} dt \: \frac{\partial}{\partial \lambda_k} \Gamma_{m,0}^* (t)
\int_0^{t_f} ds \: \Gamma_{m,0}(s)
\nonumber \\
+&
\int_0^{t_f} dt \: \Gamma_{m,0}^* (t)
\int_0^{t_f} ds \: \frac{\partial}{\partial \lambda_k} \Gamma_{m,0} (s)
\Bigg]
\nonumber \\
=
-\frac{2}{\hbar^2}
\sum_{m=1}^{\infty} 
\text{Re}
&\left[ 
\int_0^{t_f} dt \: \frac{\partial}{\partial \lambda_k} \Gamma_{m,0}^* (t)
\int_0^{t_f} ds \: \Gamma_{m,0} (s)
\right].
\end{align}
We derive an approximation to the Hessian in the same way,
\begin{align}\label{eq:hess_approx_new}
\frac{\partial^2 F }{\partial_{\lambda_l} \partial_{\lambda_k}}
\approx
-\frac{2}{\hbar^2} &\sum_{m=1}^{\infty}
\text{Re} \Big[
\nonumber \\
\int_0^{t_f} dt &\: \frac{\partial^2}{\partial \lambda_l \partial \lambda_k} \Gamma_{m,0}^* (t)
\int_0^{t_f} ds \: \Gamma_{m,0} (s)
\nonumber \\
+
\int_0^{t_f} dt &\: \frac{\partial}{\partial \lambda_k } \Gamma_{m,0}^* (t)
\int_0^{t_f} ds \: \frac{\partial}{\partial \lambda_l } \Gamma_{m,0} (s)
\Big].
\end{align}
Note that higher order terms in these approximations are obtained using \Eqref{eq:fid_expand_exact} and taking appropriate derivatives.
We highlight that \Eqref{eq:Fn} allows one in principle to calculate the fidelity approximation to any order $n$, which will improve the approximations in \Eqref{eq:eSTA_equs} that are used to construct eSTA.
One could even consider higher orders beyond the second order in \Eqref{eq:eSTA_parabola}, but this would require calculating more terms in the fidelity expansion of \Eqref{eq:fid_expand_exact} and evaluating further derivatives of \Eqref{eq:fid_expand_exact}.

To calculate $\text{eSTA}_1$ (\Eqref{eq:lambda_orig}) and $\text{eSTA}_2$ (\Eqref{eq:lambda_Hessian}) explicitly,
we write the fidelity and gradient approximations in terms of the notation from \cite{whittyQuantumControlEnhanced2020},
\begin{align}\label{eq:eSTA_fid_Gn}
F \left( \mu_s, \vec{0} \right) 
\approx 1 - 
\frac{1}{\hbar^2}
\sum_{n=1}^N \fabsq{G_n}=:f,
\end{align}
where
\begin{align}\label{eq:Gn}
G_n = \int_0^{t_f} dt \: \Gamma_{n,0}(t)\bigg|_{\mu=\mu_s,\vec{\lambda}=\vec{0}},
\end{align}
and the gradient approximation is given by
\begin{align}\label{eq:eSTA_grad_Kn}
\nabla_{\vec{\lambda}} F \left( \mu_s, \vec{0} \right) 
\approx
-\frac{2}{\hbar^2}\sum_{n=1}^N \mbox{Re} \left(G_n^* \vec{K}_n\right)
=:\vec{v},
\end{align}
where the $k^{\text{th}}$ component of $\vec{K}_n$ is given by
\begin{align}\label{eq:Kn}
\left(\vec{K}_n\right)_k = \int_0^{t_f} dt \: \frac{\partial}{\partial \lambda_k}\Gamma_{n,0} (t)\bigg|_{\mu=\mu_s,\vec{\lambda}=\vec{0}},
\end{align}
and we have truncated the infinite sums to the first $N$ terms.
We define
\begin{align}\label{eq:Wn}
\left(W_n\right)_{l,k} = 
\int_0^{t_f} dt \:
\frac{\partial^2}{\partial \lambda_l \partial \lambda_k}\Gamma_{n,0}^{*} (t)\bigg|_{\mu=\mu_s,\vec{\lambda}=\vec{0}},
\end{align}
and the entries of the Hessian approximation in \Eqref{eq:lambda_Hessian} are then given by
\begin{align}\label{eq:Hess_approx}
&\hess{ F(\mu_s, \vec{0}) }_{l,k}
=
\frac{\partial^2 F }{\partial_{\lambda_l} \partial_{\lambda_k}} \bigg\vert_{\mu=\mu_s,\vec{\lambda}=\vec{0}}
\approx
\nonumber \\
&-\frac{2}{\hbar^2} \sum_{n=1}^{N}
\text{Re} \left[ G_n \:
\left(W_n\right)_{l,k}
+
\left( \vec{K}_n^* \right)_k \: \left( \vec{K}_n \right)_l
\right]
=:\mathbb{H}_{l,k}.
\end{align}
Using \Eqref{eq:eSTA_fid_Gn}, \Eqref{eq:eSTA_grad_Kn} and \Eqref{eq:Hess_approx} we can write convenient forms for the eSTA corrections, with the improved control vector using $\text{eSTA}_1$ (\Eqref{eq:lambda_orig})
\begin{align} \label{eq:eSTA_original_final}
\vec\lambda_{s}^{(1)}  &= 2\left(1-f\right)
\frac{\vec{v} }
{\fnorm{\vec{v}}^2},
\end{align}
and for $\text{eSTA}_2$ (\Eqref{eq:lambda_Hessian}), we have
\begin{align}\label{eq:eSTA_Hessian_final}
\vec{\lambda}_{s}^{(2)} =
-\frac{\vec{v} \fnorm{\vec{v}}^2}
{\vec{v}^{\:T} \:\mathbb{H} \:\vec{v}}.
\end{align}
In the next section we apply both schemes to anharmonic trap expansion and compare the results.

%%%%%%%%%%%%%%%%%%%%%%%%%%%%%%%%%%%%%%%%%%%%%%%%%%%%%%%%%%%%%%%%%%%%%%%
\section{Anharmonic trap expansion \label{sect_trap_opening}}
%%%%%%%%%%%%%%%%%%%%%%%%%%%%%%%%%%%%%%%%%%%%%%%%%%%%%%%%%%%%%%%%%%%%%%%

We now apply eSTA to anharmonic trap expansion and compare control protocols designed using STA, $\text{eSTA}_1$ and $\text{eSTA}_2$.
Our goal is to transfer the ground state of the trap with initial trapping frequency $\omega_0$, to the ground state of the trap with final frequency $\omega_f$ with $\omega_f<\omega_0$.
We consider two trapping potentials, a Gaussian trap and an accordion lattice.
Since both potentials can be approximated by a harmonic trap near their minima, we can use a harmonic trap as the STA system from which we construct eSTA for the anharmonic systems.

\subsection{Trap potentials}\label{sect:trap_pots}
We define the accordion lattice potential as
\begin{align}\label{eq:lattice_pot}
V_L (x,t) = \text{sgn}[\omega_L(t)^2] \: A_{L} \operatorname{sin}^2\left[ k_L(t) x \right],
\end{align}
where we write the potential with $\text{sgn}[\omega_L(t)^2]$ so that negative $\omega_L(t)^2$ corresponds to the potential changing from confining to repulsive.
For the accordion lattice the wavenumber $k_L(t) = \left| \omega_L(t)\right| / \sqrt{2 A_{L}}$ is time dependent and the amplitude constant is $A_{L}=\alpha [\hbar k_L(0)]^2/2m $ with $\alpha$ a dimensionless constant fixing the size of the recoil energy.
Negative $\omega_L(t)^2$ could be implemented physically up to a global phase factor using a simple $\pi/2$ phase shift, i.e. $-A_{L} \sin^2[k_L(t)x]=A_{L}\left\{\sin^2[k_L(t)(x-\pi/2)] - 1\right\}$.

We define also the Gaussian potential as a one-dimensional approximation to an optical dipole trap \cite{luFastTransitionlessExpansions2014a}, given by
\begin{align}\label{eq:gauss_pot}
V_G (x,t) = \: A_{G}(t) \left[1-\exp \left(- k_G x^2 \right)\right].
\end{align}
The amplitude is time dependent, with $A_{G}(t) = 1/4 m w^2 \omega_G(t)^2$, $k_G = 2 \pi / \lambda_G = 2/w^2$, $m$ is the mass, $w$ the beam width and $\lambda_G$ is the trapping laser wavelength.

Using a series expansion of either potential, we have
\begin{align}\label{eq:trap_expan}
V_{0,G/L} (x,t) = \frac{1}{2} m \omega_{G/L}(t)^2 x^2 + \mathcal{O} \left(  x^{4} \right).
\end{align}
The natural choice for the STA system in the eSTA formalism for both potentials is the corresponding harmonic trap with Hamiltonian
\begin{align}\label{eq:H_0}
H_{0} = \frac{p^2}{2 m} + \frac{1}{2} m \omega_{G/L}(t)^2 x^2.
\end{align}
Note that as the trap depth is increased for both potentials they approach the limiting case of a harmonic trap.

\subsection{STA system and eSTA parametrization}
The Hamiltonian for the STA system in \Eqref{eq:H_0} has a Lewis-Leach type potential with known Lewis-Riesenfeld invariant \cite{chenFastOptimalFrictionless2010b,guery-odelinShortcutsAdiabaticityConcepts2019b}
\begin{align}\label{eq:STA_invar}
I(t)=\frac{1}{2}\left[ \frac{x^2}{b(t)^2}m\omega_0^2 + \frac{\pi(t)^2}{m} \right],
\end{align}
where $\pi = b(t)p - m \dot{b} x$ is the momentum conjugate to $x/b(t)$ and $\omega_0$ is an arbitrary constant, chosen to be $\omega(0)$ for convenience.
For $I(t)$ to be a dynamical invariant, $b(t)$ must satisfy the Ermakov equation
\begin{align}\label{eq:STA_ermakov}
\ddot{b} + \omega(t)^2 b = \frac{\omega_0^2}{b^3}.
\end{align}

We are free to choose any $b(t)$ that satisfies the appropriate boundary conditions given by $[H(t),I(t)]=0$ at $t=0,t_f$:
\begin{align}\label{eq:STA_trap_open_b}
b(0)&=1,
& \dot{b}(0)&=0,
& \ddot{b}(0)&=0,
\nonumber \\
b(t_f)&=\gamma=\sqrt{\frac{\omega_0}{\omega_f}},
& \dot{b}(t_f)&=0,
& \ddot{b}(t_f)&=0.
\end{align}
Here we use a simple polynomial ansatz for $b(t)$ from \cite{chenFastOptimalFrictionless2010b},
\begin{align}\label{eq:STA_scheme}
b(t) = 6(\gamma -1)\xi^5 - 15(\gamma - 1)\xi^4 +10 (\gamma - 1) \xi^3 + 1,
\end{align}
where $\xi=t/t_f$.
Then $\omega(t)$ can be reverse-engineered using \Eqref{eq:STA_ermakov} and we obtain  \cite{chenFastOptimalFrictionless2010b}
\begin{align}\label{eq:STA_omega} 
\omega(t)^2=\frac{\omega_0^2}{b^4} - \frac{\ddot{b}}{b}.
\end{align}
Solutions of the Sch\"odinger equation $i \hbar \partial/\partial t \Psi(x,t) = H_0 \Psi(x,t)$ can be written as,
\begin{align}\label{eq:Schr_solus}
\Psi(x,t) = \sum_{n=0}^{\infty} c_n e^{i \theta_n(t)} \psi_n(x,t),
\end{align}
where $\psi_n(x,t)$ are orthonormal eigenstates of the invariant $I$ satisfying $I(t) \psi_n(x,t)=\lambda_n \psi_n(x,t)$ and $c_n$ are constants, with the Lewis-Riesenfeld phase given by
\begin{align}
\theta_n(t) = \frac{1}{\hbar}
\int_0^t
\braXket{\phi(t',n)}{i\hbar \frac{\partial}{\partial t'}-H_0(t')}{\phi(t',n)} dt'.
\end{align}
For harmonic trap expansion, a single mode in \Eqref{eq:Schr_solus} is given by 
\begin{align}\label{eq:STA_modes}
\chi_n(x,t)&=
e^{i \theta_n(t)}
e^{i \beta_n(x,t)}
\frac{\phi_n \left( x / b\right)}{b^{1/2}}
\end{align}
where
\begin{align}
\theta_n(t) &= -(n+1/2) \int_0^t dt' \frac{\omega_0}{b(t')^2},
\nonumber \\
\beta_n(x,t) &= \frac{m}{2 \hbar}  \frac{\dot{b}}{b(t)} x^2,
\end{align}
with $\lambda_n = (n +1/2) \hbar \omega_0$, and $\phi_n(x)$ are harmonic energy eigenstates.

\begin{figure}[!t]
\begin{center}
\includegraphics[width=0.99\linewidth]{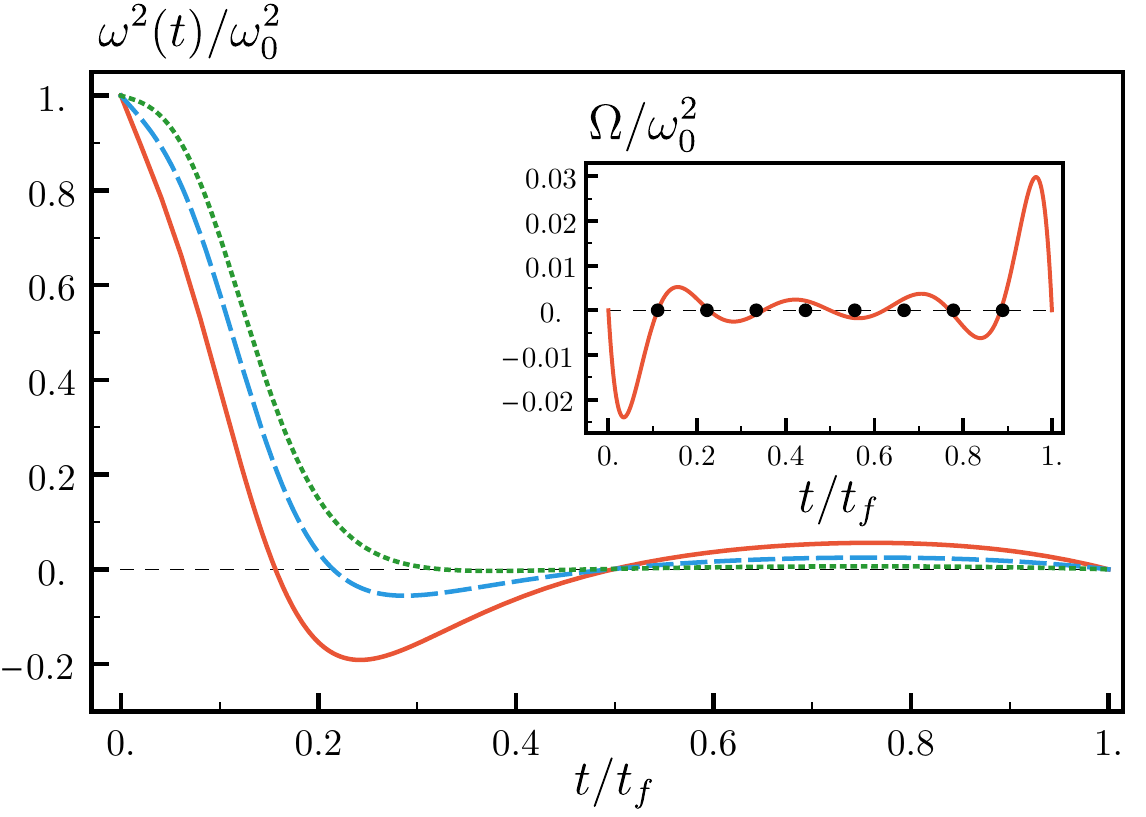}
\end{center}
\caption{\label{fig:2_trap_open_omegas}
Examples of $\omega^2(t)/\omega_0^2$ using $\gamma=10$, with $\omega_0 t_f = $ 10 (solid-red), 15 (dashed-blue), and 30 (dotted-green).
Inset: Example of $\Omega$ with $\tau_L = 26$ for fast lattice expansion using $\text{eSTA}_2$, with black dots indicating $M=8$ parameterization of \Eqref{eq:Omega_param}.
}
\end{figure}

In Fig. \ref{fig:2_trap_open_omegas}, $\omega^2(t)/\omega_0^2$ is shown for several different final times $t_f$.
Note that even though the trap frequency can become negative, there are techniques that allow negative potentials to be implemented experimentally \cite{chenFastOptimalFrictionless2010b}.
%**********************************************************************
%**********************************************************************
From \Eqref{eq:STA_omega} we obtain the STA solution $\omega(t)^2$ that we use as a starting point to construct the eSTA solution
\begin{align}
\widetilde{\omega}(t)^2 = \omega(t)^2 + \Omega (\vec{\lambda},t),
\end{align}
where for convenience we choose the eSTA correction $\Omega$ to be a polynomial that satisfies $\Omega (\vec{\lambda}, 0) = 0$ and $\Omega (\vec{\lambda},t_f) = 0$.
We parameterize $\Omega$ by the vector $\vec{\lambda}=(\lambda_1,\dots,\lambda_M)$, where
\begin{align}\label{eq:Omega_param}
\Omega\left(\vec{\lambda}, \frac{j \: t_f}{M+1}\right) = \lambda_j, \quad j=1,\dots,M,
\end{align}
and $M$ is the number of components in $\vec{\lambda}$.

Now we use eSTA to calculate the value of $\vec{\lambda}$ that improves the fidelity.
In detail, we calculate $\vec{\lambda}_s^{(1)}$ using \Eqref{eq:eSTA_original_final} ($\text{eSTA}_1$) and $\vec{\lambda}_s^{(2)}$ using \Eqref{eq:eSTA_Hessian_final} ($\text{eSTA}_2$).
These formulae require calculating  \Eqref{eq:Gn}, \Eqref{eq:Kn} and \Eqref{eq:Hess_approx}.

First we calculate $G_n$ from \Eqref{eq:Gn} using
\begin{align}
G_n &= \int_0^{t_f}dt \: \Gamma_{n,0}(t)\Big|_{\mu=\mu_s,\vec{\lambda}=\vec{0}}
\nonumber \\
&=
\int_0^{t_f}dt \: 
\braXket{\chi_n(t)}{\Delta H_{\mu_s}(\vec{0},t)}{\chi_0(t)},
\end{align}
where $\ket{\chi_n(t)}$ is given by \Eqref{eq:STA_modes} and $\Delta H_{\mu_s} = H_{\mu_s}-H_0=V_{G/L} - V_{0,G/L}$, with $H_0$ from \Eqref{eq:H_0}, $V_{G/L}$ from \Eqref{eq:gauss_pot} and \Eqref{eq:lattice_pot}, and $V_{0,G/L}$ from \Eqref{eq:trap_expan}.
To calculate the $k^\text{th}$ component of $\vec{K}_n$ from \Eqref{eq:Kn}, we have
\begin{align}
\left(\vec{K}_n\right)_k &= \int_0^{t_f}dt \: \frac{\partial}{\partial \lambda_k }\Gamma_{n,0}(t)\bigg|_{\mu_s,\vec{\lambda}=\vec{0}}
\nonumber \\
&=
\int_0^{t_f}dt \: \braXket{\chi_n(t)}{\frac{\partial}{\partial \lambda_k }\Delta H_{\mu_s}(\vec{\lambda},t)\bigg|_{\vec{\lambda}=\vec{0}}}{\chi_0(t)}
\nonumber \\
&=
\int_0^{t_f}dt \: \braXket{\chi_n(t)}{\frac{\partial}{\partial \lambda_k }V_{G/L}\bigg|_{\vec{\lambda}=\vec{0}}}{\chi_0(t)}.
\end{align}
In a similar manner we evaluate \Eqref{eq:Hess_approx} that is required only for calculating $\vec{\lambda}_s^{(2)}$.
An example of the resulting eSTA correction $\Omega(\vec{\lambda},t)$ for fast expansion of the accordion lattice using $\vec{\lambda}_s^{(2)}$ with $M=8$ components is shown in the inset of Fig \ref{fig:2_trap_open_omegas}.

\subsection{Fidelity Results: Accordion Lattice Expansion}\label{sec_fid_lattice}
We first apply eSTA to the expansion of an accordion lattice with a single trapped $^{133}$Cs atom in the ground state.
We set the lattice parameters using values from an experimentally implemented optical lattice \cite{peterDemonstrationQuantumBrachistochrones2021}, where the initial wavenumber is $k_L(0)=2\pi/ \lambda_L$ and using \Eqref{eq:trap_expan} we have that $\omega_{0,L}=\omega_L(0)=4 \sqrt{\alpha} \pi^2 \hbar / m \lambda_L^2$. 
We use numerical values $\lambda_L=866$nm and recoil energy parameter $\alpha=150$ \cite{peterDemonstrationQuantumBrachistochrones2021}.
We set the dimensionless final time $\tau_L=\omega_{0,L} t_f$ and have that $A_{L}/\hbar \omega_{0,L} \approx 6.12$.

We calculate the fidelity for different expansion times $\tau_L$.
In Fig. \ref{fig:3_fid_plots} (a) the results for STA, $\text{eSTA}_1$ and $\text{eSTA}_2$  are shown.
For both $\vec{\lambda}_{s}^{(1)}$ and $\vec{\lambda}_{s}^{(2)}$ we use  $M=1$ and $M=8$ components
($M=1$ and $M=8$ in \Eqref{eq:Omega_param} resp.).
Calculating eSTA requires truncating the sums in \Eqref{eq:Gn}, \Eqref{eq:Kn} and \Eqref{eq:Wn}.
For the results in this paper we use the first four non-zero terms.

We find that $\text{eSTA}_2$ gives improvement over $\text{eSTA}_1$ and STA as expected.
The 8 component schemes show improvement over 1 component schemes, for both $\text{eSTA}_1$ and $\text{eSTA}_2$.
This demonstrates that $\vec{\lambda}_s$ with only a few components can produce significant fidelity improvement, particularly when using $\text{eSTA}_2$.

In the derivation of $\text{eSTA}_1$ it was assumed that the system could achieve maximum fidelity, i.e. $F(\mu_s,\vec{\lambda}_{s}^{(1)})=1$.
The inset of Fig. \ref{fig:3_fid_plots} (a) demonstrates that this can lead to overshooting, which we previously illustrated schematically in Fig. \ref{fig:1_eSTA_schematic}; here we consider $\text{eSTA}_1$ and $\text{eSTA}_2$ with only 1 component, for $\tau_L = 25$.
The true fidelity landscape (solid-green), $\text{eSTA}_1$ (dotted-blue) and $\text{eSTA}_2$ (dashed-red) fidelity approximations are shown, with the $\text{eSTA}_1$ scheme minimally overshooting the optimal $\epsilon_s$.
We find that $\epsilon_{s}^{(1)}$ from $\text{eSTA}_1$ is approximately $1.5 \times \epsilon_{s}^{(2)}$.
Note that both versions of eSTA would agree if the fidelity for both $\vec{\lambda}_s^{(1)}$ and $\vec{\lambda}_s^{(2)}$ was exactly 1.
We note that calculating $\text{eSTA}_1$ may be simpler than calculating $\text{eSTA}_2$ in certain settings, and that the utility of either eSTA approach will depend on the given system dynamics.

\subsection{Fidelity Results: Gaussian Trap Expansion}\label{sec_fid_gaussian}
We consider Gaussian trap expansion and use similar values to the Gaussian approximation of a single trapped $^{87}$Rb atom in an optical dipole trap in \cite{luFastTransitionlessExpansions2014a,torronteguiFastTransitionlessExpansion2012}, with inverse unit of time $\omega_{0,G}=2\pi \times 2500$Hz, $ A_{G}/\hbar \omega_{0,G} \approx 2418$, laser wavelength $\lambda_{\text{laser}}=1060$nm, a beam waist of $20\lambda_{\text{laser}}$ and set the expansion time $\tau_G=\omega_{0,G} t_f$. 

\begin{figure}[!ht]
(a) \includegraphics[width=0.99\linewidth]{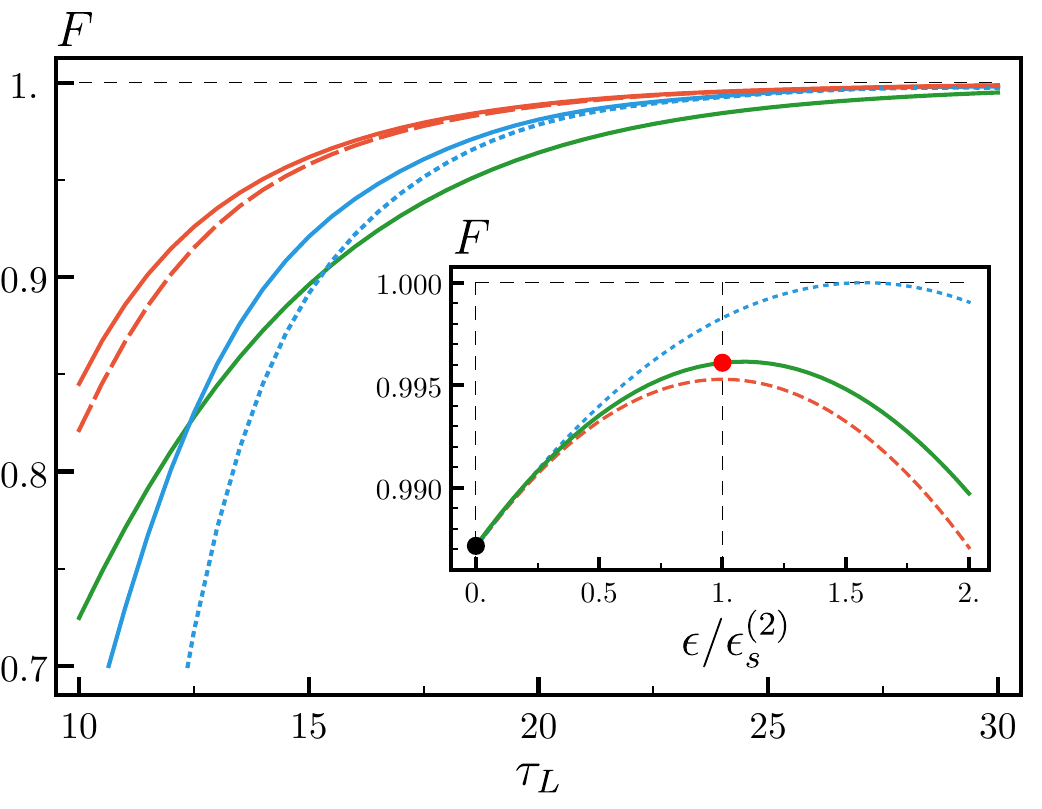}\\
(b) \includegraphics[width=0.99\linewidth]{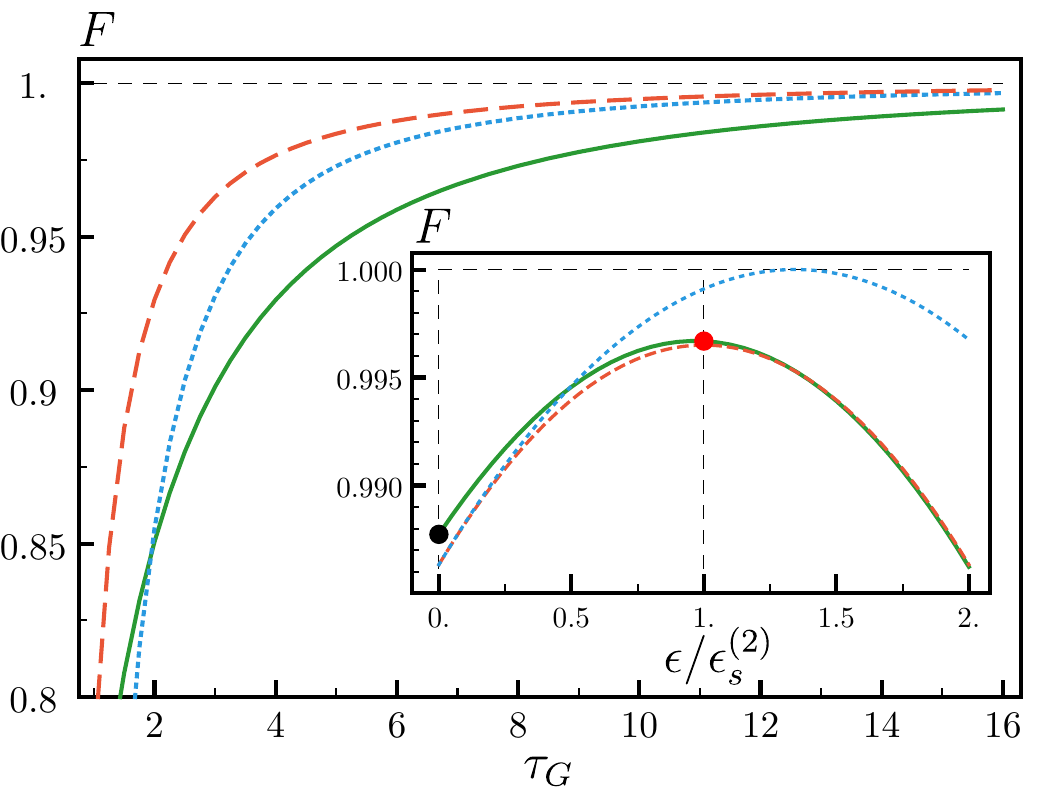}
\caption{Fidelity vs. expansion time $\tau_{L/G}$.
(a) Lattice expansion; STA (solid green), 
$\text{eSTA}_1$ with $M=8$ components (solid blue) and $M=1$ (dotted blue),
$\text{eSTA}_2$ $M=8$ (solid red) and $M=1$ (dashed red).
Lattice parameters as in Sec. \ref{sec_fid_lattice} for this plot and the inset.
\newline
Inset: Fidelity vs $\epsilon/\epsilon_s^{(2)}$ for lattice expansion with $\tau_L=25$; true fidelity landscape (solid green), $\text{eSTA}_1$ (dotted blue) and $\text{eSTA}_2$ (dashed red) parabola approximations.
\newline
(b) Gaussian trap expansion; same labeling as (a), with $M=1$ and $M=8$ results indistinguishable (solid lines omitted).
Physical values given in Sec. \ref{sec_fid_gaussian}. Inset: Same labeling as (a) with $\tau_G = 13$.
}\label{fig:3_fid_plots}
\end{figure}

We simulate Gaussian trap expansion for different expansion times $\tau_G$ and the results are shown in Fig. \ref{fig:3_fid_plots} (b), using the STA scheme (solid-green line), $\text{eSTA}_1$ (dotted and solid blue lines) and $\text{eSTA}_2$ (dashed and solid red lines). 

As with the optical accordion, we consider $\text{eSTA}_1$ and $\text{eSTA}_2$ with two parameterizations of $\vec{\lambda}_{s}^{(1)}$ and $\vec{\lambda}_{s}^{(2)}$, a 1 component scheme (dotted-blue, dashed-red) and an 8 component scheme (solid blue, solid red).
We find that $\text{eSTA}_1$ and $\text{eSTA}_2$ are an improvement over STA, and that the 1 and 8 component schemes produce very similar results for both $\text{eSTA}_1$ and $\text{eSTA}_2$.
This is an indication that the polynomial form of $\Omega(\vec{\lambda},t)$ allows a large class of improved schemes.
The inset of Fig. \ref{fig:3_fid_plots} (b) demonstrates again that the original eSTA scheme can minimally overshoot the optimal $\epsilon_s$ (compare again to Fig. \ref{fig:1_eSTA_schematic}), with the parabola calculated using $\text{eSTA}_2$ matching the true fidelity well.

%%%%%%%%%%%%%%%%%%%%%%%%%%%%%%%%%%%%%%%%%%
%%%%%%%%%%%%%%%%%%%%%%%%%%%%%%%%%%%%%%%%%%

\subsection{ESTA Sensitivity}
In this section we consider errors in the trapping potentials and calculate the sensitivity to these errors using the STA, $\text{eSTA}_1$ and $\text{eSTA}_2$ schemes introduced earlier.
For the lattice potential we consider an error in the amplitude
\begin{align}
V_{\text{err}}^L(x,t) = \text{sgn}[\omega_L(t)^2] \: A_{L}(1+\delta) \operatorname{sin}^2\left[ k_L(t) x \right],
\end{align}
and for the Gaussian potential we also consider an amplitude error, given by
\begin{align}
V_{\text{err}}^G(x,t) = A_{G}(1+\delta) \left\{1 - \operatorname{exp}^2 \left[ -k_G(t) x^2 \right]\right\}.
\end{align}
We define the error sensitivity by
\begin{align}\label{eq:S_systematic}
S := \fabs{\frac{\partial F}{\partial \delta}\Bigr|_{\delta=0}},
\end{align}
and calculate this quantity numerically using a multi-point discrete approximation to the derivative.
Note that a lower sensitivity $S$ means a given scheme is more robust against error.

Heuristically we expect that eSTA will simultaneously improve fidelity and robustness: for $\mu=0$ both eSTA and STA give fidelity $1$, and as $\mu$ increases the eSTA fidelity is always higher than the STA fidelity, i.e. the slope of the eSTA fidelity is expected to be smaller than the slope of the STA scheme. 
Assuming that this slope is approximately proportional to the error sensitivity $S$, we also expect lower error sensitivity for eSTA than STA.

\begin{figure}[t]
\begin{center}
(a) \includegraphics[width=0.99\linewidth]{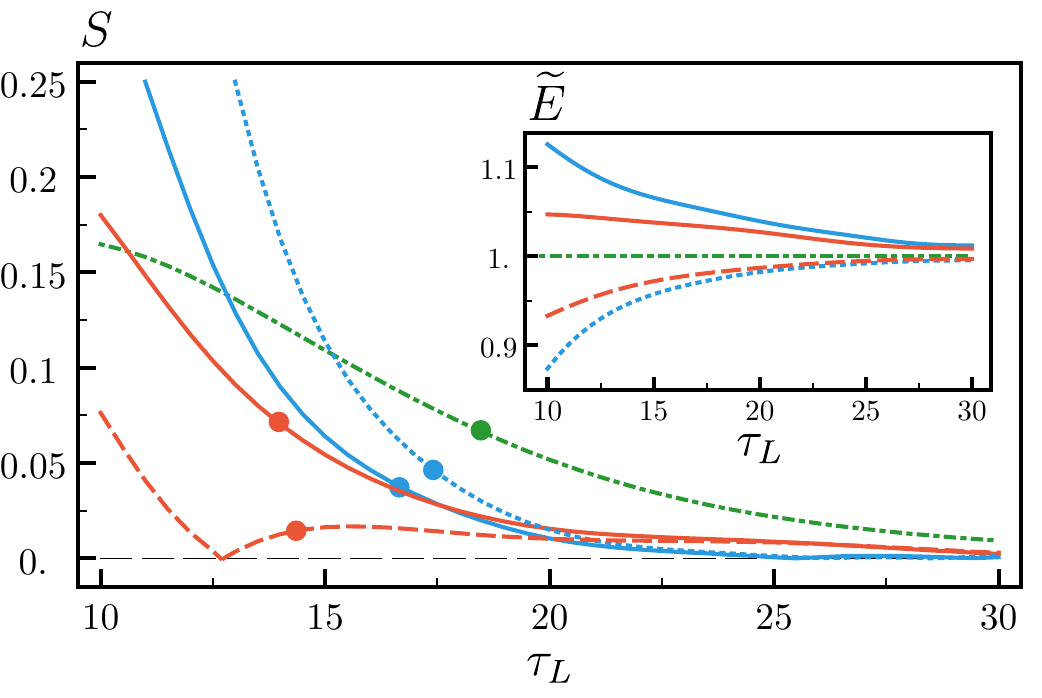} \\
(b) \includegraphics[width=0.99\linewidth]{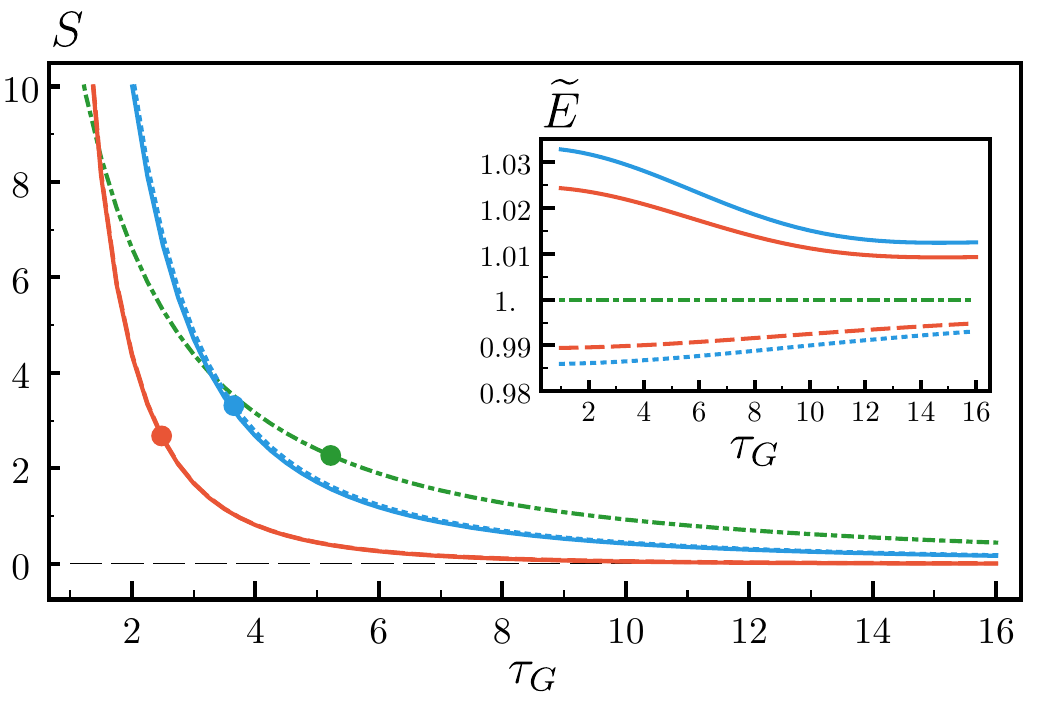}
\caption{Trap expansion sensitivity $S=\fabs{\partial F / \partial \delta}$ vs. $\tau_{L/G}$ Same parameters and labeling as Fig \ref{fig:3_fid_plots}.
(a) Lattice expansion, with first $\tau_L$ for which $F>0.95$ marked on each line. Inset: Time averaged energy (\Eqref{eq:time_avg_E}) of the eSTA schemes scaled by the STA scheme (dot-dashed green), $\widetilde{E}$ vs. $\tau_L$.
(b) Gaussian trap expansion sensitivity $S$ vs. $ \tau_G$, again with first $\tau_G$ for which $F>0.95$ shown.
\label{fig:4_sens_plots}}
\end{center}
\end{figure}

In the following we look at the numerical error sensitivity.
In Fig \ref{fig:4_sens_plots} (a) the sensitivity of lattice expansion is shown, with STA (dot-dashed green), $\text{eSTA}_1$ ($M=1$ dotted-blue, $M=8$ solid blue) and $\text{eSTA}_2$ ($M=1$ dashed-red, $M=8$ solid red).
Each line is marked at the point where $F \ge 0.95$, with $\text{eSTA}_1$ and $\text{eSTA}_2$ still achieving this fidelity for lower $\tau_L$ than STA.
In this high fidelity regime, both eSTA schemes are generally less sensitive (smaller $S$) to error than the STA scheme for $\tau_L \gtrsim 15$, in agreement with the heuristic argument from above.
The $\text{eSTA}_2$ scheme generally gives the highest fidelities and lowest sensitivities, as shown in Fig. \ref{fig:3_fid_plots} (a) and Fig. \ref{fig:4_sens_plots} (a).
Interestingly, the single component ($M=1$) $\text{eSTA}_2$ scheme (dashed-red line) has lower sensitivity than the 8 component ($M=8$) $\text{eSTA}_2$ scheme (solid-red line).

%%%%%%%%%%%%%%%%%%%%%%%%%%%%%%%%%%%%%%%%%%
%%%%%%%%%%%%%%%%%%%%%%%%%%%%%%%%%%%%%%%%%%

For Gaussian trap expansion, see Fig. \ref{fig:4_sens_plots} (b), there is negligible difference in sensitivity between choosing a single or 8 component scheme, for either $\text{eSTA}_1$ or $\text{eSTA}_2$.
Again, the points for which $F \ge 0.95$ is first achieved are marked on each line.
For these high fidelities $\text{eSTA}_1$ and $\text{eSTA}_2$ outperform STA, in agreement with the heuristic argument outlined earlier.
In this case $\text{eSTA}_2$ has generally the highest fidelity, as shown in Fig. \ref{fig:3_fid_plots} (b), as well as the lowest sensitivity in Fig. \ref{fig:4_sens_plots} (b).
A convenient eSTA scheme can be chosen depending on the required fidelity or sensitivity. 

We also consider the time averaged energy
\begin{align}\label{eq:time_avg_E}
E (\vec{\lambda})= \frac{1}{t_f}\int_0^{t_f} dt \:
\langle H_{\mu_s}(\vec{\lambda},t) \rangle,
\end{align}
and define $\widetilde{E} =  E (\vec{\lambda}) / E (\vec{0})$ such that the different eSTA protocols are in direct comparison with the STA scheme.
The insets in Fig. \ref{fig:3_fid_plots} (a) and (b) show $\widetilde{E}$ for $\text{eSTA}_1$ and $\text{eSTA}_2$.
The values of $\widetilde{E}$ of the eSTA schemes are close to $1$ (i.e. close to the STA scheme), demonstrating that little additional time averaged energy is required for improvement.
In both lattice and Gaussian expansion, the 1 component STA schemes have a lower time averaged energy than STA ($\widetilde{E}<1$), while using more components ($M=8)$ has a higher value ($\widetilde{E}>1$).

%%%%%%%%%%%%%%%%%%%%%%%%%%%%%%%%%%%%%%%%%%%%%%%%%%%%%%%%%%%%%%%%%%%%%%%
\section{Conclusion}\label{sect_conclusion}
%%%%%%%%%%%%%%%%%%%%%%%%%%%%%%%%%%%%%%%%%%%%%%%%%%%%%%%%%%%%%%%%%%%%%%%

The main result in this paper is a generalization of the original eSTA derivation in \cite{whittyQuantumControlEnhanced2020}, and the construction of an alternative $\text{eSTA}$ scheme.
This alternative eSTA scheme allows the removal of an assumption of the original eSTA method, at the expense of calculating an additional Hessian term.
Both eSTA schemes are applied to anharmonic trap expansion, resulting in higher fidelity and improved robustness.

Generally, there are several important advantages that the eSTA formalism has to offer; the derivation is analytic, applicable to a wide variety of quantum control problems and the control schemes are expected to have enhanced robustness against noise.
In addition, eSTA can offer insight into a given control problem , for example by first considering low dimension parameterizations of the control scheme.
There is also significant freedom in choosing how to parameterize the control scheme for either approach;
for example, we can choose to preserve the symmetry of the original STA scheme, or use a form of the eSTA improvement that lends itself to certain conditions e.g. a Fourier sum with fixed bandwidth.
Analytic eSTA control schemes that are outside the class of STA solutions can be derived, and they could give improved starting points for numerical optimization.
As an outlook, higher order eSTA schemes can be constructed using the formalism presented in this paper which would be useful if some lower order terms vanish.

\begin{acknowledgments}
We are grateful to D. Rea and J. Li for their fruitful discussion and careful reading of the manuscript.
C.W. acknowledges support from the Irish Research Council (GOIPG/2017/1846).
A.K. acknowledges support from the Science Foundation Ireland Starting Investigator Research Grant ``SpeedDemon'' No. 18/SIRG/5508.
A.R. acknowledges support from the Science Foundation Ireland Frontiers for the Future Research Grant ``Shortcut-Enhanced Quantum Thermodynamics'' No. 19/FFP/6951.
\end{acknowledgments}

\bibliography{eSTA_trap_expansion}{}
\bibliographystyle{apsrev4-1}

\end{document}